\documentclass[10pt,conference]{IEEEtran}
\usepackage{bm}
\usepackage{cite}
\usepackage{mathrsfs}
\usepackage{textcomp}
\usepackage{epsfig}
\usepackage{indentfirst}
\usepackage{amsmath}
\DeclareMathOperator{\tr}{tr}
\usepackage{amssymb}
\usepackage{algorithm}
\usepackage{algorithmicx}
\usepackage{algpseudocode}
\usepackage{multirow}
\usepackage{float}
\usepackage{array}
\usepackage{multirow}
\usepackage{graphicx}
\usepackage{subfigure}
\usepackage{amsthm}
\theoremstyle{definition}

\theoremstyle{remark}

\usepackage{cite}
\begin{document}
\title{A Hybrid Approach for Efficient Wireless Information and Power Transfer in Green C-RAN}

\author{
	\IEEEauthorblockN{
		{Xu Li, Zhao Chen, Aurobinda Laha, Ziru Chen, Yu Cheng and Lin X. Cai}
		\IEEEauthorblockA{
			Department of Electrical and Computer Engineering,
			Illinois Institute of Technology, Chicago, USA\\
			Email: xli230@hawk.iit.edu, zchen84@iit.edu, alaha@hawk.iit.edu, zchen71@hawk.iit.edu, \{cheng, lincai\}@iit.edu
		}
	}
	\thanks{This work is supported in part by NSF grants CNS-1320736, ECCS-1610874, NSF Career award ECCS1554576, and National Natural Science Foundation of China (NSFC) under grant 61628107.}
}

\maketitle

\begin{abstract}
In this paper, we consider a green cloud radio access network (C-RAN) with simultaneous wireless and power transfer ability. In order to reduce the energy consumed for updating the channel state information (CSI), energy users are divided into two different groups, including the free charge group and the MIMO group. Then a semi-definite programming problem is formulated under the constraints of energy and information transmission requirements. To minimize the total energy consumption, two algorithms are developed to authorize the energy users into two group divisions in single time slot. Then the algorithms are extended to long term scenarios consisting training and long term stages, the CSI of free charge energy users are not required during the long term stage. Simulation and numerical results are presented to demonstrate the efficiency of the proposed algorithms in significantly reducing the energy consumption of C-RAN systems.
\end{abstract}


\section{Introduction}

% Green Cloud-RAN

In next generation wireless communications, cloud radio access network (C-RAN) has been proposed~\cite{chih2014toward} as a promising network architecture that
incorporates cloud computing into wireless mobile networks, which can not only mitigate inter-cell interference, but also greatly reduce the capital expenditure (CAPEX) and operational expenditure (OPEX) cost.
In C-RAN systems, remote radio heads (RRHs) will be densely deployed, which are connected to a baseband unit (BBU) pool via a high-capacity low-latency fronthaul link. In order to reduce the grid power consumption and make the C-RAN system more environment friendly, green C-RANs powered by smart-grid technologies have been investigated~\cite{sparse}, where the RRHs are supplied with renewable energy sources such as solar and wind powers and energy trading and sharing between nearby RRHs are also supported.

% wireless charging, energy harvesting and SWIPT

Meanwhile, it is worth noting that the radio-frequency (RF) energy of wireless signals can be also exploited to charge mobile users' batteries~\cite{cai2}.
In the literature, it has been demonstrated that ambient wireless signals~\cite{vyas2013wehp} and dedicated energy signals~\cite{ju2014throughput} can be both utilized to charge low-power devices.
Particularly, simultaneous wireless information and power transfer (SWIPT)~\cite{xu2014multiuser} can complete data transmission and RF energy charging at the same time.

% CSI in energy beamforming

Based on these observations, to enhance the C-RAN systems with SWIPT capability~\cite{sparse,chen2017optimal,ng2015secure} will offer the capability to support many new functions and better user experience with longer battery life in the realm of IoT and mobile networks.
To support SWIPT in C-RANs, networked MIMO beamforming is widely adopted, where the requirement of channel state information will highly impact on the system energy efficiency, especially for C-RANs.
In \cite{sparse}, the author proposed algorithms to divide the energy terminals (ETs) into two different groups where less energy is offered to ETs that are far from the RRHs. In \cite{Dynamic}, the authors proposed a method to achieve higher energy efficiency in wireless power transfer (WPT) by using instantaneous CSI. \cite{PARTCSI} introduced a design for beamforming and minimum mean-square-error (MMSE) based on partly known CSI of channel. In \cite{xu2014multiuser}, a joint information and energy transmit beamforming design was formulated to maximize the weighted sum-power received by energy users.

% In \cite{caimesh}, network with supply of sustainable energy and its charging and discharging process are discussed. \cite{cai2} discusses a fully sustainable cooperative system with multiple sources nodes.

% However, the low energy efficiency rate of wireless power transfer affect the popularity of the technology. The fading characteristics of the wireless signal and quality of service (QoS) demand for both data and energy users increase the operational expenses extensively.

From the above mentioned prior works, we notice that either full or partial CSI is required for all the ETs and information terminals (ITs), which will consume extra energy for users on channel estimation and CSI feedback. Especially for devices powered by RF energy, such energy consumption is critical and will definitely shorten their life time.
% However, for the ETs equipped with a battery, it is sufficient to satisfy them with long-term average received RF energy.
% Thus, we can use the average channel to analyze the energy harvesting of ETs in long term situation.
However, we observe that for the users that are sufficiently close to the RRHs, the average received RF energy in long-term transmissions is much larger than their requirement even without MIMO beamforming.
Therefore, we propose a novel hybrid approach for wireless information and power transfer, which eliminates the CSI requirement for some ETs located within a elaborately optimized range of RRHs.
Such selected ETs are referred to as free charge users in this paper.
The main contributions of this paper are summarized as follows:
\begin{itemize}
\item We propose an energy-efficient green C-RAN system to support SWIPT using a hybrid approach. Specifically, some ETs can be freely charged without reporting their CSI to save energy and prolong their life time.
\item To determine ET group division with instantaneous CSI, we introduce two range based iterative algorithms, where the group division will be updated according to the optimization problem which minimizes the total system energy consumption while ensuring the QoS demand for both ITs and ETs.
\item To optimize ET group for long-term transmissions, we proposed a training based algorithm depending on the proposed iterative algorithms, which has been verified to save energy significantly with extensive simulations.
\end{itemize}

\section{System model}

\begin{figure}[t]
	\centering
	\includegraphics[width=2.5in]{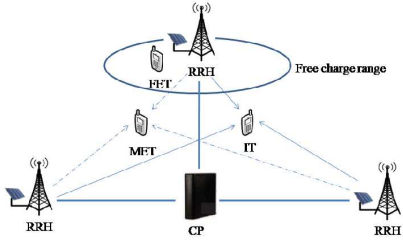}
	\caption{System model with optimized free Charge range.}
	\label{fig:system model}
\end{figure}

As shown in Fig. \ref{fig:system model}, we consider a green C-RAN system with SWIPT capability in this paper, which consists of $N$ single-antenna RRHs, $U_E$ single-antenna ETs and $U_D$ single-antenna ITs.
All RRHs are connected to one central processor (CP), which is responsible for resource allocation and coordinates the joint beamforming at each RRH.
Let $\mathcal{L}_D = \{1,2,\ldots,U_D\}$, $\mathcal{L}_E = \{1,2,\ldots,U_E\}$ and $\mathcal{N} = \{1,2,\ldots,N\}$ represent the index sets of ITs, ETs and RRHs, respectively.
In our model, ETs are dynamically divided into two independent groups, namely free charge ETs (FETs) and MIMO ETs (METs).
The relationship between any two ET groups is given as:
\begin{align}
\mathcal{L}_E & =\mathcal{L}_M \cup \mathcal{L}_F, \mathcal{L}_M \cap \mathcal{L}_F = \varnothing
\end{align}
where we define $\mathcal{L}_M = \{1,2,\ldots,L_M\}$, $\mathcal{L}_F = \{L_M+1,L_M+2,\ldots, U_E\}$.
The ETs harvest energy and the ITs receive the information from the RRHs at the same time by exploiting the downlink channel.

% while the METs and ITs simultaneously feed their CSI back via the lossless uplink channel. In this paper, we assume the downlink and uplink are perfectly isolated with each other. The basic system model is shown in Fig.\ref{fig:system model}.

% As the energy-carrying signal contains no information but the power to offer the energy transfer and the ET can harvest energy from information signals due to the broadcast property of the downlink channel, we assume that the system only offer the direct data beamforming towards ITs while it coordinates the energy terminals sharing the same beamforming power via the same band frequency in the downlink channel.

Let $\boldsymbol{\omega}_i = [\omega_{1i}^H,\ldots,\omega_{Ni}^H]^H \in \mathbb{C}^{N \times 1}$ represent the beamforming vector from all RRHs towards the $i$-th IT, $\forall i \in \mathcal{L}_D$, where $\omega_{ni}$ denotes the beamforming weight from $n$-th RRH towards the $i$-th IT. Let $\boldsymbol{h}_i^{[ID]} = [(h_{1i}^{[ID]})^H,\ldots,(h_{Ni}^{[ID]})^H]^H \in \mathbb{C}^{N \times 1}$ represent the channel vector  from all RRHs towards the $i$-th IT, $\forall i \in \mathcal{L}_D$. Let $\boldsymbol{h}_i^{[MET]} = [(h_{1i}^{[MET]})^H,\ldots,(h_{Ni}^{[MET]})^H]^H \in \mathbb{C}^{N \times 1}$ represent the channel vector from all RRHs towards the $i$-th MET, $\forall i \in \mathcal{L}_E$. Hence, the received information-carrying signal $y_i$ at the $i$-th IT can be written as:
\begin{align}
y_i & = (\boldsymbol{h}_{i}^{[ID]})^H{\boldsymbol{\omega}}_{i}s_{i} + \sum_{j \in \mathcal{L}_D,j \neq i}{(\boldsymbol{h}}_{i}^{[ID]})^H{\boldsymbol{\omega}}_{j}s_{j}+ n_{i},
\end{align}
The received signal consists of three parts: the faded desired information-carrying signal $s_i$ via the Rayleigh fading channel, the interference signal $s_j$ that is supposed to be delivered from RRHs to other ITs due to the sharing bandwidth, and the Gaussian noise $n_i \sim \mathcal{CN}(0,\sigma^2_i)$ at $i$-th IT side. Without loss of generality, we have assumed that $\mathbb{E}(s_i) = 1, \forall i \in \mathcal{L}_D$ for the rest of the paper.

To measure the QoS for the ITs, we can derive the signal-to-interference-plus-noise-ratio (SINR) of the $i$-th IT as:
\begin{align}
SINR_i=\frac{|({\boldsymbol{h}}_{i}^{[ID]})^H{\boldsymbol{\omega}}_{i}|^2}{\sigma^2_i + \sum_{j \in \mathcal{L}_D,j \neq i}|({\boldsymbol{h}}_{i}^{[ID]})^H{\boldsymbol{\omega}}_{j}|^2}
\end{align}
In our model, different energy transmission service is performed based on the ETs' group division, i.e., METs are considered to receive the energy via beamforming transmission, in which extra energy is required to reporting CSI to CP via uplink channel for real time beamforming coordination. FETs are considered to receive the energy via the information signal broadcast of single RRH, due to the fading of the signal, one RRH can only guarantee free charge service in a short distance, this distance will be named as free charge range in the rest of our paper, hence the energy that are carried by signals from other RRHs has been ignored.

We assume that the number of ETs is fixed and limited. In each time slot, under the perfect CSI knowledge of METs and ITs, The beamforming vector for joint transmission is optimized by CP. Hence, the total energy harvested by $i$-th MET, $\forall i \in \mathcal{L}_M$ in each time slot is expressed as:
\begin{align}
P_i^{[M]} & =\eta\sum_{j \in \mathcal{L}_D}|({\boldsymbol{h}}_{i}^{[MET]})^H{\boldsymbol{\omega}}_{j}|^2
\end{align}
where $\eta$ represents energy harvesting efficiency rate. In this paper, we assume that all the ETs have the same energy harvesting efficiency.

The energy that FETs can harvest in long term situation is considered as the average energy that it received from the broadcasting signal delivered by the closest RRH, as each ET is equipped with a battery to storage energy. The closest RRH of one FET will be considered as the assigned RRH for this FET. The distance $D_i^{[F]}$ from $i$-th free charge user, $\forall i \in \mathcal{L}_F$, to its assigned RRH is given by:
\begin{align}
D_i^{[F]} & = \min_{\forall n \in \mathcal{N}} D_{in}^{[F]}
\end{align}
where $D_{in}^{[F]}$ represents the distance from $i$-th FET to the $n$-th RRH. Hence, the average energy that a single free charge energy user can harvest in each slot from its assigned RRH is:
\begin{align}
P^{[F]}_{i} & = \eta P^{[op]}_n(D_i^{[F]})^{\frac{1}{\alpha}},
\end{align}
where $P_n^{[op]}$ is the total energy consumption used for signal delivery of each RRH. The minimum energy that an FET requires is denoted as $P_{fmin}$, and the minimum energy that an MET requires is $P_{amin}$. Although the METs and FETs are essentially same type of devices, due to consumption of updating CSI, the minimum energy that is required by METs is larger than FETs, as $P_{fmin} < P_{amin}$.

We assume that the total energy cost will not reach the maximum operational power limitations of each RRH. Hence the $P_n^{[op]}$ of each RRH is:
\begin{align}
P_n^{[op]} & = \sum_{i \in \mathcal{L}_D}|\omega_{n i}|^2, \forall n \in \mathcal{N}.
\end{align}
The CP coordinates the beamforming based on the CSI updated by MIMO terminals in long term situation. In order to guarantee the minimum received energy requirement of all FETs, there exists the free charge range set $\mathcal{RAN}$ , in which all the ETs can be served as FETs. Thus the free charge service range $Range$ of $n$-th RRH is determined as below:
\begin{align}
Range_n & =  \left(\eta P_n^{[op]}/P_{fmin}\right)^{\frac{1}{\alpha}},
\end{align}
where $\alpha$ is the path loss exponent of this system. The green energy that is generated by devices installed in each RRHs is considered as $P_n^{[en]},\forall n \in \mathcal{N}$. So we can calculate the total energy cost for $n$-th RRH $P^{[pu]}_n$ as:
\begin{align}
P^{[pu]}_n = P_n^{[op]} - P_n^{[en]}.
\end{align}
Notice that the $P_n^{[pu]}$ for each RRH must be non-negative due to the system characteristics.

\section{Problem Formulation}
In this paper, we aim to minimize the total system energy cost to satisfy the energy harvest requirement of ETs and the QoS requirement of ITs by adjust the beamforming vector of each RRH. We consider the total energy cost of each RRH in this system as $P_n^{[pu]}$ and the total energy that is used for information delivery and energy transmission defined as $P_n^{[op]}$. Now, we can formulate the following optimization problem:
\begin{align}
\min_{{P^{[op]}},P^{[pu]}} \hspace{1em} & \beta\sum_{n \in \mathcal{N}}P^{[pu]}_n + \gamma\sum_{n \in \mathcal{N}}P^{[op]}_n, \forall n \in \mathcal{N}\\
s.t. \hspace{1em}  & SINR_i \geq SINR_{min}, \forall i \in \mathcal{L}_D\\
                   & P_{i}^{[M]} \geq P_{amin},\forall i \in \mathcal{L}_M \\
                   & P_i^{[F]} \geq P_{fmin}, \forall i \in \mathcal{L}_F\\
                   & P^{[op]}_n \leq P^{[pu]}_n + P_n^{[en]}, \forall n \in \mathcal{N}\\
                   & \forall P^{[pu]}_n \geq 0,\forall n \in \mathcal{N}
\end{align}
where $SINR_{min}$ is the minimum SINR that is required by all ITs to guarantee the QoS constraint. The weight coefficients $\beta$ and $\gamma$ both larger than 0 is set to balance the system and avoid multi-solution problem in the convex optimization.

Let $\boldsymbol{H}_i =\boldsymbol{h}^H_i \boldsymbol{h}_i$ and $\boldsymbol{W}_i = \boldsymbol{\omega}^H_i \boldsymbol{\omega}_i$. $\boldsymbol{D}_n \triangleq diag(0_0 \ldots 0_n,1,0_{n+1} \ldots 0_N)\succeq 0,\forall n \in \mathcal{N} $ denote an auxiliary matrix. $\sum_{n \in \mathcal{N}}P^{[op]}_n = \sum_{i \in \mathcal{L}_D}\sum_{n \in \mathcal{N}}\boldsymbol{\omega}^H_i \boldsymbol{\omega}_i\boldsymbol{D}_n = \sum_{i \in \mathcal{L}_D}\boldsymbol{W}_i$ represent the total energy that is used to deliver the signals. Let $\boldsymbol{D}^{mf}_i$ represents the auxiliary $\boldsymbol{D}$ matrix that is  associated with the $i$-th, $\forall i \in \mathcal{L}_F$ free charge user. The constraint of SINR for $i$-th, $\forall i \in \mathcal{L}_D$ IT user can be written as:

%0_1 \ldots 0_i,\ldots,1_n \ldots 0_n
\begin{align}
\frac{\tr({\boldsymbol{H}_i}{\boldsymbol{W}_i})}{\sum_{j \in \mathcal{L}_D, j \neq i}({\boldsymbol{H}_i}{\boldsymbol{W}_j}) + \sigma^2_i} \geq SINR_{min}
\end{align}
Hence, the problem in (10) can be written as
\begin{align}
\min_{P^{[pu]},\boldsymbol{W}} \hspace{1em} & \beta\sum_{n \in \mathcal{N}}P^{[pu]}_n + \gamma\sum_{i \in \mathcal{L}_D}\boldsymbol{W}_i, \forall n \in \mathcal{N}, \forall i \in \mathcal{L}_D\\
s.t. \hspace{2em}  & \frac{\tr({\boldsymbol{H}_i}{\boldsymbol{W}_i})}{SINR_{min}} - \sum_{j \in \mathcal{L}_D, j \neq i}\tr({\boldsymbol{H}_i}{\boldsymbol{W}_j}) - \sigma^2_i \geq 0,\nonumber\\ &\forall i \in \mathcal{L}_D\\
                   & \sum_{i \in \mathcal{L}_D}{\tr({\boldsymbol{H}}}_{j}{\boldsymbol{W}}_{i}) \geq P_{amin},\forall j \in \mathcal{L}_M\\
                   & \frac{\eta \sum_{\forall i \in \mathcal{L}_D}{\tr{(\boldsymbol{W}_i}{\boldsymbol{D}^{[mf]}_j})}}{(d^{[mf]}_j)^{\alpha}} \geq P_{fmin}, \forall j \in \mathcal{L}_F\\
                   & \sum_{\forall i \in \mathcal{L}_D}\tr{(\boldsymbol{W}_i}{\boldsymbol{D}_n}) \leq P^{[pu]}_n + P_n^{[en]},\forall n \in \mathcal{N}\\
                   & P^{[pu]}_n \geq 0,\forall n \in \mathcal{N}
\end{align}
When $P^{[en]}_n - P^{[op]}_n \geq 0$ is true for all the $n$ in $\mathcal{N}$ and both weight coefficients are large than 0, the problem (17) can be transformed into
\begin{align}
\min_{P^{[pu]},\boldsymbol{W}_{i}} \hspace{1em} &{(\beta + \gamma)}*{\sum_{i \in \mathcal{L}_D}\boldsymbol{W}_i}-\gamma *P^{[en]}_n
\end{align}
If $P^{[en]}_n - P^{[op]}_n \leq 0, \forall n \in \mathcal{N}$, due to the constraint in (22) as the system can only receive the energy from the grid, the energy cost in this RRH is considered to be 0, In order to avoid the multi-solution problem and reduce the energy consumption. We choose the solution which have minimum energy cost in each RRH as the optimal result.

\section{Range Based Optimal Group Division}
In this paper, we propose two iterative algorithms to derive the optimal ET division authorization and generate beamforming design based on the ET division. As the system constraints in the optimization problem (17), the set of FETs $\mathcal{L}_F$ and METs $\mathcal{L}_M$ is dynamic during the iterations, so we can get the optimal result of problem (17) based on the ET group division of each iteration round. In this section, we discuss the situation of single time slot, then compare the difference between the two proposed algorithms and evaluate their performance. Finally, we extend them to long term scenarios.

\subsection{Algorithms for single slot situation}

The key component of both algorithms is based on the free charge range of each RRH decided by the system settings and the ET group division of the current iteration round. The acceptable free charge range of each RRH can be computed as:
\begin{align}
  Range_n = \left[\frac{\eta \sum_{i \in \mathcal{L}_D}{\tr{\boldsymbol({W}_i}{\boldsymbol{D}_{n}})}} {P_{fmin}}\right]^{\frac{1}{\alpha}}, \forall n \in \mathcal{N}
\end{align}
As $\mathcal{RAN}$ represent the set of free charge range of all RRHs, and if the distance of one ET to its assigned RRH is closer than the free charge range of this RRH, this particular terminal will be authorized as an FET in the iterative algorithm.

We set $\boldsymbol{Group}$ as the ET group division including $\mathcal{L}_F$ and $\mathcal{L}_M$ and $\boldsymbol{Group}_k$ as the group division after $k$-th iteration round respectively. Hence, we define the process that is used for updating the ET group division in each algorithm.

\floatname{algorithm}{\allowdisplaybreaks}
\renewcommand{\algorithmicrequire}{\textbf{Input:}}
\renewcommand{\algorithmicensure}{\textbf{output:}}
\begin{algorithm}
        \caption{Group division updating process}
        \begin{algorithmic}[1]
            \State Input the system setting, ET group division $\boldsymbol{Group}_{k-1}$
            \State Calculate problem (17) for optimal RRH energy cost $P^{[Pu]}$ and beamforming vector set $\boldsymbol{W}$
            \State Calculate the free charge range $\mathcal{RAN}_k$ based on the $\boldsymbol{W}$ and system setting
            \State Updating the ET Group division $\boldsymbol{Group}_{k}$ based on the system setting and free charge range $\mathcal{RAN}_k$
            \State Check the boundary point
  \end{algorithmic}
\end{algorithm}

The system settings include the locations of each RRH, ET and IT, the CSI of each channel, $P_{amin}, P_{fmin}, SINR_{min}$ and other system settings that was initially present. This process has also been refered as one single iteration round in this paper. It can be noticed that the system settings remain fixed once initialized. In the iteration process, there may exist some terminals which are located within short distance from the free charge range boundary but we consider those points as boundary terminal. For boundary points, we will compare the system performance when setting this point as FET or MET and choose the better result as output.

To adjust the different situations, we developed two different algorithms based on the group division updating process. In the first algorithm, we begin with the initial ET group division with $\boldsymbol{L}_{M} = \boldsymbol{L}_{E}, \boldsymbol{L}_{M} = \boldsymbol{Group}_{0}$, which means that all the ETs will be considered as METs in the beginning. The algorithm is described below.

\floatname{algorithm}{\allowdisplaybreaks}
\renewcommand{\algorithmicrequire}{\textbf{Input:}}
\renewcommand{\algorithmicensure}{\textbf{output:}}
\begin{algorithm}
        \caption{Algorithm 1:Iterative algorithm considering METs}
        \begin{algorithmic}[1] %Ã¿ÐÐÏÔÊ¾ÐÐºÅ
            \State Input the system settings and $\boldsymbol{Group}_{0}$
            \State Process channel checking, generate $\boldsymbol{Group}_{1}$
            \State Process Group division update and generate $\boldsymbol{Group}_{2}$
            \While {$(\boldsymbol{Group}_{k-1} \neq \boldsymbol{Group}_{k},\forall k \geq 0)$}
               \State Process Group division update
            \EndWhile
            \If {$(\boldsymbol{Group}_{k-1} = \boldsymbol{Group}_{k})$}
            \State Choose $P^{[pu]}$ set generated by $\boldsymbol{Group}_{k-1}$ as output
            \Else
            \State Compare all the system output results generated by $\boldsymbol{Group}_{(k-n)}$ to $\boldsymbol{Group}_{k}$, when $\boldsymbol{Group}_{(k-n)} = \boldsymbol{Group}_{k}$, Choose the minimum $P^{[pu]}$ set as output
            \EndIf
            \State Total system energy consumption $P^{[total]} = \sum_{n \in \boldsymbol{N}}P_n^{[pu]}$
  \end{algorithmic}
\end{algorithm}

In the initial part, when the energy terminal channel towards its assigned RRHs is extremely poor, the burden of energy transmission to support this user will be  separately taken by other surrounding RRHs whose channel have better fading rate but longer distance. This will affect the iteration processes as this energy terminal will be considered as an MET with high rate energy cost. In order to avoid this, we consider this energy terminal as Free charge energy terminal in the beginning round of iteration, an action which we name as 'channel check process'. In order to avoid the endless iteration loop that may exist in some situations, we consider that if current ET group division is equal to any division in the past iterations, then we break the loop and take the minimum system energy cost result between the current group and the duplicate group in past as the output.

In the next proposed algorithm, we consider the $\boldsymbol{Group}_{0}$ decided by the free charge range that is generated by the green energy as the RRH have no battery to store the extra energy. The initial range of each RRH can be expressed as:
\begin{align}
  Range_n = [\eta P^{[en]}_n/ P_{fmin}]^{\frac{1}{\alpha}}, \forall n \in \mathcal{N}
\end{align}
So in this algorithm, ETs inside of this range will be considered as FETs and other ETs will be considered as METs before the channel check process. The algorithm is described below.

\floatname{algorithm}{\allowdisplaybreaks}
\renewcommand{\algorithmicrequire}{\textbf{Input:}}
\renewcommand{\algorithmicensure}{\textbf{output:}}
\begin{algorithm}[H]
        \caption{Algorithm 2:Iterative algorithm with green energy}
        \begin{algorithmic}[1] %Ã¿ÐÐÏÔÊ¾ÐÐºÅ
            \State Input the system settings
            \State Process the initial range $\mathcal{RAN}_{0}, \forall n \in \boldsymbol{N}$ based on the green energy harvested
            \State Decide the group division $\boldsymbol{Group}_0$ based on $\boldsymbol{Ran}_{0}$ and the system settings
            \State Process the channel checking, generate $\boldsymbol{Group}_1$
            \State Rest parts same as algorithm 1 start from step 3
  \end{algorithmic}
\end{algorithm}

The major advantage of this algorithm compared to the Algorithm 1 is that the system will not require the CSI for the ET that in the green energy free charge range. However, if there are no ETs inside that range, it will take one more iteration round to achieve the result as the second iteration round is same as the initial round of Algorithm 1.

\subsection{Algorithm for long term scenarios}

For long term scenarios, we divide the long term in to two stages, the training stage and the long term stage. In the training stage, we assume that all ETs are updating the CSI to CP in each time slot regardless which ET group division they are in and initialize the system to process the iteration in each time slot. After $Q^{training}$ slots, we can get the possibility of one ET being an FET or MET and then we can get ET group division training results under a certain threshold.

In the long term stage, the ET group division remains fixed and equal to the training result we get in the training stage, the METs and ITs will keep updating the CSI in each time slot while FETs remain silent in the uplink channel.

%\begin{figure}[H]
%	\centering
%	\includegraphics[width = 3.2in]{location.png}
%	\caption{RRHs and terminals location.}
%	\label{fig:RRH}
%\end{figure}
\begin{figure*} [t]
%  \centering
%  \hfill
  %
  \begin{minipage}{0.32\textwidth}
%    \centering
    \includegraphics[width=2.5in]{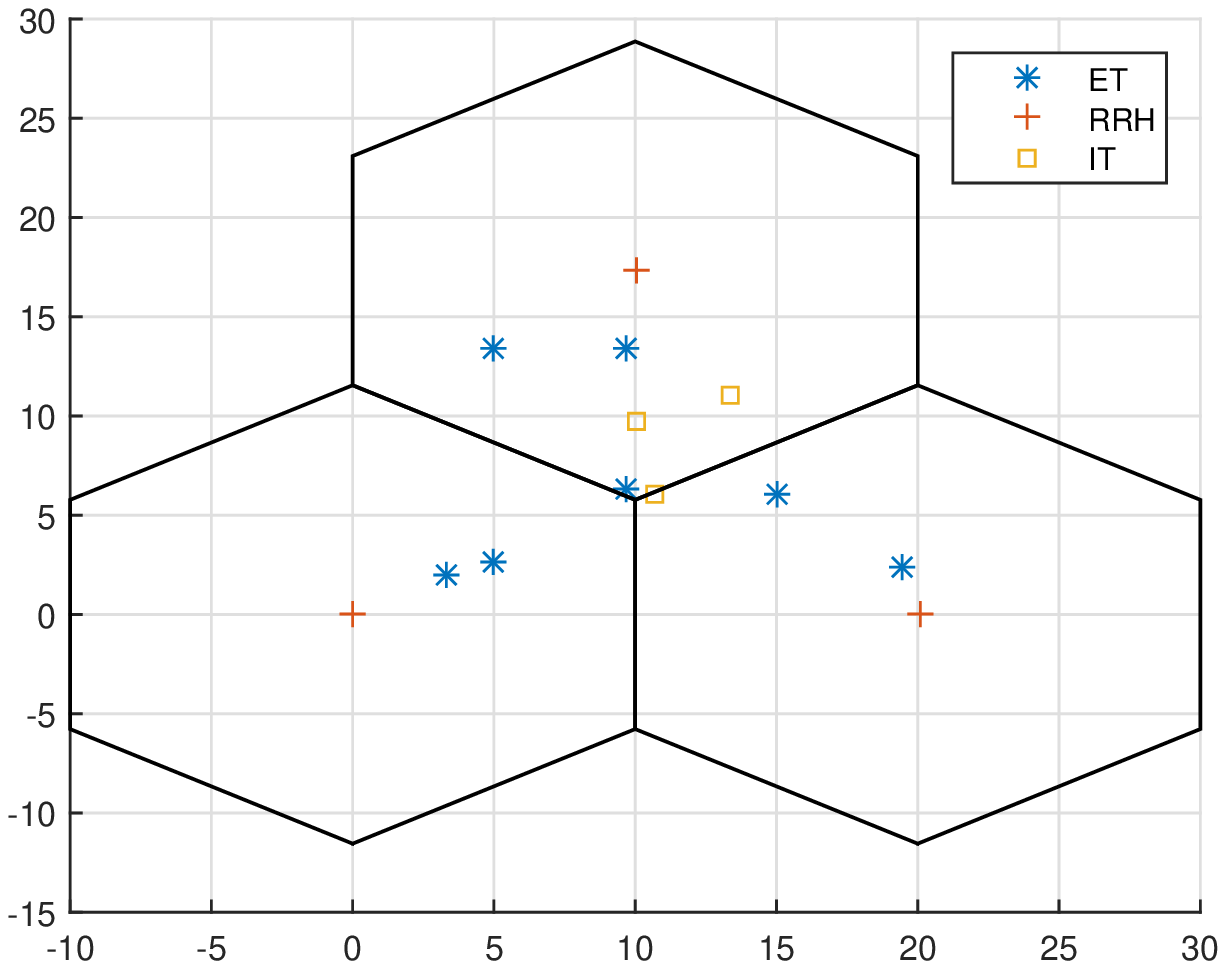}
    \caption{RRHs and terminals location.}\label{fig:RRH}
  \end{minipage} %
 % \hfill
 %
 \begin{minipage}{0.32\textwidth}
%    \centering
    \includegraphics[width=2.5in]{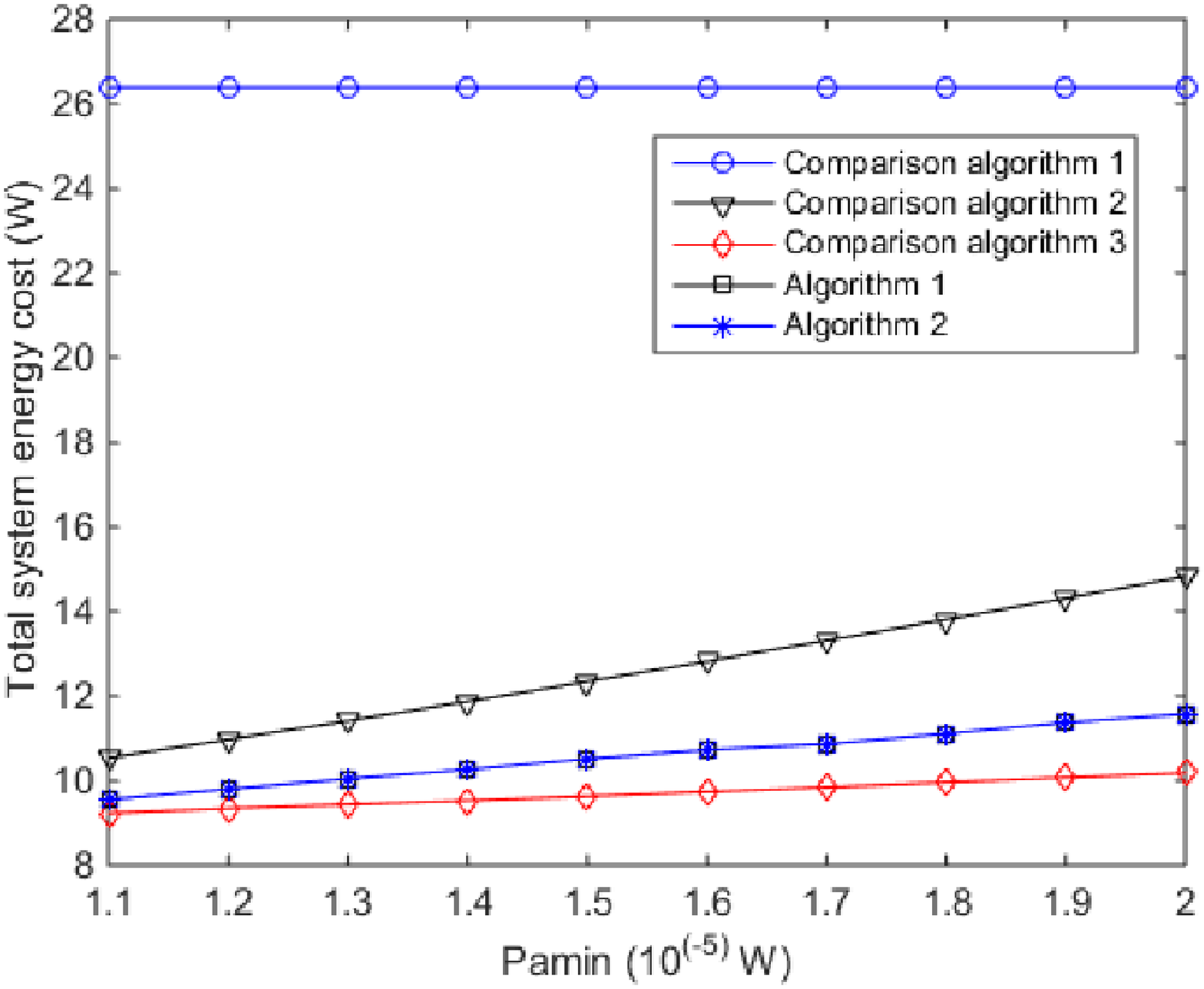}
 %   \caption{Bonding probability $P_{CB}$ (2 channels)}
    \caption{System performance of total energy consumption under different $P_{amin}$ in single time slot situation.}
  \label{fig:pamin}
  \end{minipage}
%  \hfill
 %
 \begin{minipage}{0.32\textwidth}
%  	\centering
  \includegraphics[width=2.5in]{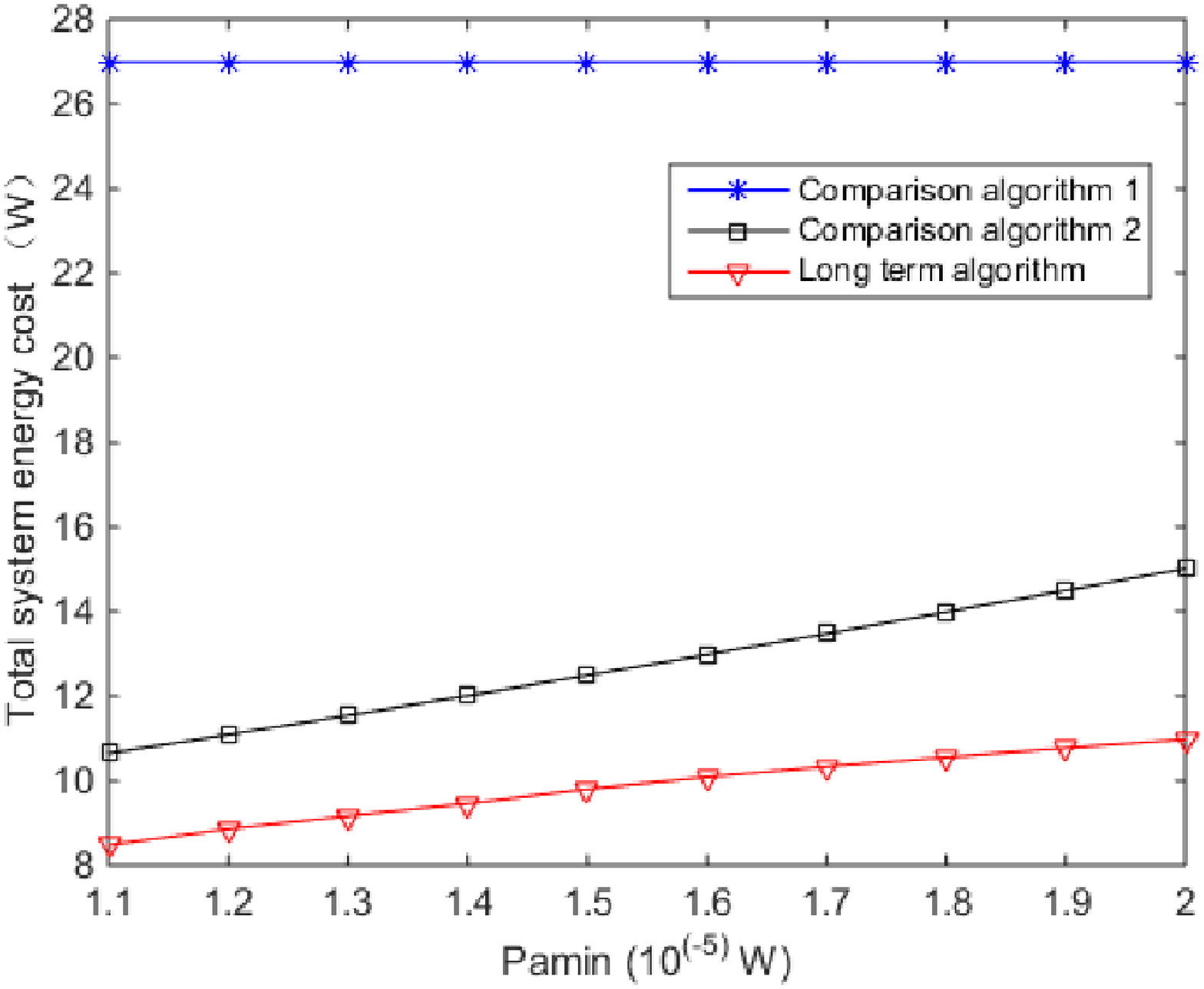}
    \caption{System performance of total energy consumption under long term scenarios.} \label{fig:multiple}
   \end{minipage}
%  \hfill
%

\end{figure*}
\section{Simulation Results and Analysis}
We set up a SWIPT system with 3 RRHs, 7 ETs and 4 ITs where each RRH is located with in 20m distance from each other in neighboring hexagonal cells. ETs and ITs locations are randomly generated in three RRH hexagonal cells as Fig.\ref{fig:RRH}. The simulation parameters are set as follow: The green energy $P^{[en]}$ that harvested by each RRH is 2, 2.5, 3 W respectively. The $SINR_{min}$ required by ITs is set as 20 and the minimum energy required by FETs and METs is -20 dBm and -17 dBm. The path loss exponent $\alpha$ is -2.5 and energy harvesting efficiency for all ETs is 80$\%$. The energy division threshold is 0.5, the weight coefficients $\beta$ and $\gamma$ are set as 1.

%The green energy $P^{[en]}$ generated by each renewable device is 2, 2.5, 3W. The minimum SINR that is required by ITs is 20 and the minimum energy $P_{fmin}$ and $P_{amin}$ required by ETs is -20 dbm and -17 dbm. The path loss exponent $\alpha$ is -2.5, The energy harvest efficiency $\eta$ is 0.8 for all ETs. The weight factors $\beta$ and $\gamma$ are equal to 1. The ET energy division threshold is 0.5 .%
The simulation of each case is operated with CVX under SDP mode for the output of average over 200 individual channel sets. We also set up several comparison algorithms to analyze our performance. In the first comparison algorithm we consider that all ET in this system are FETs, the second comparison algorithm is that we consider all ETs in this system are METs which is commonly used in MIMO settings of current research. The third comparison algorithm is the brute force algorithm where we test all the possible ET group divisions under this channel set and choose the lowest system cost as the output. This output can be considered as the optimal value that this system can achieve with current system setting.

%\begin{figure}[ht]
% \centering
% \includegraphics[width = 3.2in]{thesispamin.png}
% \caption{System performance of total energy consumption under% different $P_{amin}$ in single time slot situation.}
% \label{fig:pamin}
%\end{figure}

Fig.\ref{fig:pamin} shows the system perform under different $P_{amin}$. When $P_{amin}$ is increasing, the situation is equal to the energy that used for updating CSI is increasing. As the $P_{amin}$ increase, the free charge range is increasing to support the energy requirement of METs, but as our algorithms can dynamically adjust the ET group division base on the range, so the it have a better performance compare to the comparison algorithm 2 when CSI consumption is large.

%\begin{figure}[H]
% \centering
% \includegraphics[width = 3.2in]{thesislongrun.png}
% \caption{System performance of total energy consumption under long% term scenarios.}
% \label{fig:multiple}
%\end{figure}

In the long term scenarios, we set the number of training slots to be 10, and take one training stage and one long term stage as a complete sample, the comparison algorithms 1 and 2 are the same as the single slot situation, the performance of this system is shown in Fig.\ref{fig:multiple}.

\section{Conclusion}
In this paper, we have focused on energy saving research by using the free charge service to support the energy transmission to some ETs based on the beamforming design. ETs are dynamically divided into two groups, then an optimization problem is formulated to minimize the total energy consumption with guaranteeing the QoS requirements for both ITs and ETs. Next, two iterative algorithms are proposed to authorize the ET group division based on the free charge range and then an extended algorithm in long term scenarios is presented. The performance evaluation of proposed algorithms in single slot and long term scenarios are presented which show that compared to the current SWIPT green C-RAN models, our algorithms are able to significantly reduce the energy consumption.

\section*{Acknowledgement}
This work is supported in part by NSF grants CNS-1320736, ECCS-1610874, NSF Career award ECCS1554576, and National Natural Science Foundation of China (NSFC) under grant 61628107.
% references section

% can use a bibliography generated by BibTeX as a .bbl file
% BibTeX documentation can be easily obtained at:
% http://mirror.ctan.org/biblio/bibtex/contrib/doc/
% The IEEEtran BibTeX style support page is at:
% http://www.michaelshell.org/tex/ieeetran/bibtex/
\bibliographystyle{IEEEtran}
% argument is your BibTeX string definitions and bibliography database(s)

\bibliography{IEEEabrv,reference}

% Generated by IEEEtran.bst, version: 1.14 (2015/08/26)
\begin{thebibliography}{10}
\providecommand{\url}[1]{#1}
\csname url@samestyle\endcsname
\providecommand{\newblock}{\relax}
\providecommand{\bibinfo}[2]{#2}
\providecommand{\BIBentrySTDinterwordspacing}{\spaceskip=0pt\relax}
\providecommand{\BIBentryALTinterwordstretchfactor}{4}
\providecommand{\BIBentryALTinterwordspacing}{\spaceskip=\fontdimen2\font plus
\BIBentryALTinterwordstretchfactor\fontdimen3\font minus
  \fontdimen4\font\relax}
\providecommand{\BIBforeignlanguage}[2]{{%
\expandafter\ifx\csname l@#1\endcsname\relax
\typeout{** WARNING: IEEEtran.bst: No hyphenation pattern has been}%
\typeout{** loaded for the language `#1'. Using the pattern for}%
\typeout{** the default language instead.}%
\else
\language=\csname l@#1\endcsname
\fi
#2}}
\providecommand{\BIBdecl}{\relax}
\BIBdecl

\bibitem{chih2014toward}
I.~Chih-Lin, C.~Rowell, S.~Han, Z.~Xu, G.~Li, and Z.~Pan, ``Toward green and
  soft: a {5G} perspective,'' \emph{IEEE Commun. Mag.}, vol.~52, no.~2, pp.
  66--73, 2014.

\bibitem{sparse}
W.~N. S. F.~W. Ariffin, X.~Zhang, and M.~R. Nakhai, ``Sparse beamforming for
  real-time resource management and energy trading in green c-ran,'' \emph{IEEE
  Transactions on Smart Grid}, vol.~8, no.~4, pp. 2022--2031, July 2017.

\bibitem{cai2}
Z.~Chen, L.~X. Cai, Y.~Cheng, and H.~Shan, ``Sustainable cooperative
  communication in wireless powered networks with energy harvesting relay,''
  \emph{IEEE Transactions on Wireless Communications}, vol.~PP, no.~99, pp.
  1--1, 2017.

\bibitem{vyas2013wehp}
R.~J. Vyas, B.~B. Cook, Y.~Kawahara, and M.~M. Tentzeris, ``{E-WEHP}: A
  batteryless embedded sensor-platform wirelessly powered from ambient
  digital-{TV} signals,'' \emph{IEEE Trans. Microw. Theory Techn.}, vol.~61,
  no.~6, pp. 2491--2505, 2013.

\bibitem{ju2014throughput}
H.~Ju and R.~Zhang, ``Throughput maximization in wireless powered communication
  networks,'' \emph{IEEE Trans. Commun.}, vol.~13, no.~1, pp. 418--428, 2014.

\bibitem{xu2014multiuser}
J.~Xu, L.~Liu, and R.~Zhang, ``Multiuser {MISO} beamforming for simultaneous
  wireless information and power transfer,'' \emph{IEEE Trans. Signal
  Process.}, vol.~62, no.~18, pp. 4798--4810, 2014.

\bibitem{chen2017optimal}
Z.~Chen, Z.~Chen, L.~X. Cai, and Y.~Cheng, ``Optimal beamforming design for
  simultaneous wireless information and power transfer in sustainable
  cloud-ran,'' \emph{IEEE Transactions on Green Communications and Networking},
  2017.

\bibitem{ng2015secure}
D.~W.~K. Ng and R.~Schober, ``Secure and green {SWIPT} in distributed antenna
  networks with limited backhaul capacity,'' \emph{IEEE Trans. Wireless
  Commun.}, vol.~14, no.~9, pp. 5082--5097, 2015.

\bibitem{Dynamic}
G.~Yang, C.~K. Ho, and Y.~L. Guan, ``Dynamic resource allocation for
  multiple-antenna wireless power transfer,'' \emph{IEEE Transactions on Signal
  Processing}, vol.~62, no.~14, pp. 3565--3577, July 2014.

\bibitem{PARTCSI}
C.~Xing, N.~Wang, J.~Ni, Z.~Fei, and J.~Kuang, ``Mimo beamforming designs with
  partial csi under energy harvesting constraints,'' \emph{IEEE Signal
  Processing Letters}, vol.~20, no.~4, pp. 363--366, April 2013.

\end{thebibliography}
%
% <OR> manually copy in the resultant .bbl file
% set second argument of \begin to the number of references
% (used to reserve space for the reference number labels box)

\end{document}